\documentclass[twocolumn,showpacs]{revtex4}
\usepackage{amsthm,amssymb,amsmath}

\newcommand\opk[1]{\mathop{\mathrm{#1}}\nolimits}
\newcommand\tr{\opk{tr}}
\newcommand\rk{\opk{rk}}
\newcommand\spn{\opk{span}}
\newcommand\abs[1]{\left|#1\right|}

\newcommand\ket[1]{| #1 \rangle}
\newcommand\ketbra[1]{| #1 \rangle\langle #1 |}

\newcommand\Ga{\Gamma}
\newcommand\al{\alpha}
\newcommand\om{\omega}

\newcommand\Hil{\mathcal{H}}

\newtheorem{theorem}{Theorem}

\newtheorem{corollary}{Corollary}

\theoremstyle{remark}
\newtheorem{example}{Example}
\newcommand\defn[1]{\textsl{#1}}

\begin{document}

\title{Entanglement of subspaces in terms of entanglement of superpositions}
\author{Gilad Gour}
\email{gour@math.ucalgary.ca}
\author{Aidan Roy}
\email{aroy@qis.ucalgary.ca}
\affiliation{Institute
for Quantum Information Science and Department of Mathematics and Statistics, 
University of Calgary, 2500 University Drive NW, Calgary AB T2N 1N4, Canada}

\date{\today}

\begin{abstract}
We investigate upper and lower bounds on the entropy of entanglement of a superposition of bipartite states as a function of the individual states in the superposition. In particular, we extend the results in [G. Gour, arxiv.org:0704.1521 (2007)] to superpositions of several states rather than just two. We then investigate the entanglement in a subspace as a function of its basis states: we find upper bounds for the largest entanglement in a subspace and demonstrate that no such lower bound for the smallest entanglement exists. Finally, we consider entanglement of superpositions using measures of entanglement other than the entropy of entanglement.
\end{abstract}
\maketitle

\section{Introduction}
\label{sec:intro}

In recent years entanglement has been recognized as 
the key resource for many quantum information processing tasks such as teleportation
and super-dense coding.  This recognition led to an intensive search for mathematical tools
that would enable a proper quantification of this resource. For pure states, the entropy of entanglement has been found to be the unique measure of entanglement in the asymptotic limit of many copies. This remarkable property can not be extended to mixed states and despite the enormous effort in recent years, mixed bipartite entanglement is far from being completely understood~\cite{Hor07,Ple06}. 

One of the difficulties in the quantification of mixed entanglement is linked to the fact the entanglement of a superposition of pure bipartite states can not be simply expressed as a function of the entanglement of the individual states in the superposition. This is because entanglement mostly depends on the coherence among the states in the superposition. It is therefore somewhat surprising that there exist tight lower and upper bounds on the entanglement of a superposition of two states in terms of the entanglement of the individual states in the superposition. In this paper we extend the results given in~\cite{Gour07} to include superpositions of more than two states and use this extension to investigate the entanglement presented in bipartite subspaces, or in short, the entanglement of subspaces~\cite{GourNolan07}. More specifically, given a subspace spanned by an orthonormal bipartite basis, we ask how the states with minimum and maximum entanglement depend on the entanglement of the individual states in the basis. While several 
authors~\cite{Nolan02,Hayden06,Cubitt07,Walgate07} 
have recently investigated the entanglement of a subspace as a function of its dimension, entanglement as a function of a basis is less well studied. 

Our interest in the minimum entanglement of a subspace derives from its connection to a well-known conjecture in quantum information: the additivity of quantum channel output entropy. Recall that a quantum channel $N$ is a completely positive trace-preserving linear map, and may be characterized as having the form
\[
N(\rho) := \tr_B(U \rho U^\dag),
\]
for some unitary matrix $U$ and some subsystem $B$. The minimum entropy output is the minimum value of $S(N(\rho))$ over all density matrices $\rho$. Since entropy is a concave function and $N$ is linear, it suffices to consider density matrices of pure states, $\rho = \ketbra{\phi}$, with $\ket{\phi}$ in some input space $\Hil^I$. Letting $V$ denote the range of $U$, say $V = \spn\{U\ket{\phi}: \ket{\phi} \in \Hil^I\}$, and assuming $V$ is a subspace of some bipartition $\Hil^A \otimes \Hil^B$, we have
\[
\min_{\ket{\phi} \in \Hil^I} S(N(\ketbra{\phi})) = \min_{\ket{\psi} \in V} E(\psi).
\]
Therefore, finding the minimum entropy output of $N$ is equivalent to finding the minium entropy of entanglement in $V$.

The well-known conjecture is that minimum entropy output is additive:
\[
\min_{\rho} E(N_1 \otimes N_2(\rho)) = \min_{\rho} E(N_1(\rho)) + \min_{\rho} E(N_2(\rho)).
\]
It follows that subspace entanglement is additive if and only if minimum entropy output for quantum channels is additive. By the results of Shor \cite{Shor04}, additivity of subspace entanglement is therefore also equivalent to additivity of the Holevo capacity of a quantum channel, additivity of entanglement of formation, and strong superadditivity of entanglement of formation. Moreover, any lower bound for subspace entanglement is also a bound for minimum entropy output. For example, in~\cite[Proposition 2]{GourNolan07} a lower bound 
is given for the entanglement of a subspace $U \otimes V$ in terms of the entanglment of $U$ and $V$ and the dimension $d = \min\{\dim U, \dim V\}$:
\[
\min_{\ket{\psi} \in U \otimes V} E(\psi) \geq \min_{\ket{\psi} \in U} E(\psi) + \min_{\ket{\psi} \in V} E(\psi) - \log d.
\]
In terms of minimum entropy output, their bound says that 
\[
\min_{\rho} E(N_1 \otimes N_2(\rho)) \geq \min_{\rho} E(N_1(\rho)) + \min_{\rho} E(N_2(\rho)) - \log d,
\]
where $d$ is smaller of the dimensions of the ranges of $N_1$ and $N_2$.
 
The outline of the paper is as follows. In Section \ref{sec:states}, we consider bounds for the entanglement of a superposition in terms of the entanglement of its components. There are both upper (Theorems \ref{thm:uppergenstate} and \ref{thm:upperinductionstate}) and lower (Theorem~\ref{thm:lowerbound}) bounds. In Section \ref{sec:spaces}, we use the bounds in Section \ref{sec:states} to search for bounds on the maximum and minimum entanglement that occurs in a subspace. Finally, in Section \ref{sec:measures}, we point out how the techniques used in this paper can be applied to measures of entanglement other than the entanglement of formation.

\section{Entanglement for superposition of more than two states}
\label{sec:states}

In this section, we derive bounds on the entanglement of a superposition of states, as a function of the entanglement of the components. More precisely, if $\ket{\psi}$ is a bipartite state shared by $A$ and $B$, then the entanglement of $\ket{\psi}$ is measured by the von Neumann entropy of the reduced state of either party:
\[
E(\psi_i): = S(\tr_A\ketbra{\psi}) = S(\tr_B\ketbra{\psi}).
\]
Given $\ket{\Ga} = \sum_{i=1}^n \al_i \ket{\psi_i}$, we find bounds on $E(\Ga)$ in terms of $E(\psi_i)$. Bounds for the superposition of two states were first considered by Linden, Popescu, and Smolin~\cite{Linden06}; their results were improved and generalized by Gour~\cite{Gour07}. Our results generalize Gour's to superpositions of more than two states.

First consider upper bounds. For superpositions of two states, the result~\cite[Theorem 3]{Gour07} is the following:

\begin{theorem}
\label{thm:uppertwostate}
Let $\ket{\psi_1}$ and $\ket{\psi_2}$ be normalized, orthogonal bipartite states, and let $\ket{\Ga} = \al_1\ket{\psi_1} + \al_2\ket{\psi_2}$, also normalized. Then for any $0 \leq p \leq 1$,
\[
E(\Ga) \leq \left(\frac{\abs{\al_1}^2}{p} + \frac{\abs{\al_2}^2}{1-p}\right)\!\Big(p E(\psi_1) + (1-p) E(\psi_2) + H(p)\Big).
\]
\end{theorem}

For a superposition of several states, Theorem \ref{thm:uppertwostate} can be generalized as follows:

\begin{theorem}
\label{thm:uppergenstate}
Let $\{ \ket{\psi_i} \}_{i=1}^n$ be a set of normalized, orthogonal bipartite states, 
and let $\ket{\Ga} = \sum_i \al_i\ket{\psi_i}$, also normalized. Then for any $\{p_i\}$ such that $p_i \geq 0$ and $\sum_i p_i = 1$, 
\[
E(\Ga) \leq \left(\sum_i \frac{\abs{\al_i}^2}{p_i}\right)\Big(\sum_i p_i E(\psi_i) + H(\{p_i\})\Big).
\]
\end{theorem}

\begin{proof}
Let $\ket{\chi_1} = \ket{\Ga}$, and suppose we choose $\ket{\chi_2}, \ldots, \ket{\chi_n}$ so that 
\begin{equation}
\sum_i q_i \ketbra{\chi_i} = \rho = \sum_i p_i \ketbra{\psi_i}.
\label{eqn:unittrans}
\end{equation}
for some $\{\ket{\chi_i}\}$. Then the entropy of $\rho$ is bound above and below by the entropies of $\{\chi_i\}$ and $\{\psi_i\}$:
\[
\sum_i q_i E(\chi_i) \leq S(\tr_A \rho) \leq \sum_i p_i E(\psi_i) + H(\{p_i\}).
\]
In particular,
\begin{equation}
q_1 E(\Ga) \leq \sum_i p_i E(\psi_i) + H(\{p_i\}).
\label{eqn:gabound}
\end{equation}
It remains to find the value of $q_1$ as a function of $\al_i$ and $p_i$. Equation~\eqref{eqn:unittrans} holds if and only if $\{\sqrt{q_i}\ket{\chi_i}\}$ and $\{\sqrt{p_i}\ket{\psi_i}\}$ are related by some unitary transformation $U$. Let $(a_i)$ denote the first row of the transformation $U$, so that 
\[
\sqrt{q_1} \ket{\Ga} = \sum_i a_i\sqrt{p_i} \ket{\psi_i}.
\]
But we also have $\ket{\Ga} = \sum_i \al_i\ket{\psi_i}$, so matching coefficients we get $\abs{a_i}^2 = \abs{\al_i}^2q_1/p_i$. Now
\[
1 = \sum_i \abs{a_i}^2 = q_1 \sum_i \frac{\abs{\al_i}^2}{p_i},
\]
so $1/q_1 = \sum_i \abs{\al_i}^2/p_i$, and the result follows from equation~\eqref{eqn:gabound}.
\end{proof}

There is another way to bound from above the entanglement of a superposition of $n$ states: using a superposition of two states recursively. We begin by simplifying and weakening Theorem \ref{thm:uppertwostate}. Noting that $H(p) \leq 1$ and then minimizing over all $p$, Theorem \ref{thm:uppertwostate} implies the following:

\begin{corollary}
\label{cor:uppertwostate}
Let $\ket{\psi_1}$ and $\ket{\psi_2}$ be normalized, orthogonal bipartite states, and let $\ket{\Ga} = \al_1\ket{\psi_1} + \al_2\ket{\psi_2}$, also normalized. Then 
\[
E(\Ga) \leq \left(\abs{\al_1}\sqrt{E(\psi_1)+1} + \abs{\al_2}\sqrt{E(\psi_2)+1}\right)^2.
\]
\end{corollary}

We can now use Corollary \ref{cor:uppertwostate} inductively two obtain a bound on the superposition of $n$ states.

\begin{theorem}
\label{thm:upperinductionstate}
Let $\{ \ket{\psi_i} \}_{i=1}^n$ be a set of normalized, orthogonal bipartite states, and let $\ket{\Ga} = \sum_i \al_i\ket{\psi_i}$, also normalized. Then
\[
E(\Ga) \leq \Big(\sum_{i=1}^n \abs{\al_i}\sqrt{E(\psi_i) + 1}\Big)^2 + n-2.
\]
\end{theorem}

\begin{proof}
For convenience, set 
\[
G_i := \abs{\al_i}\sqrt{E(\psi_i) + 1},
\]
so that we are trying to prove $E(\Ga) \leq (\sum_i G_i)^2 + n-2$. By Corollary \ref{cor:uppertwostate}, the result is true in the case $n=2$. Now assume by way of induction that the result is true for $n-1$. Let $\ket{\Phi} := \sum_{i=1}^{n-1} \frac{\al_i}{\sqrt{1-\abs{\al_n}^2}} \ket{\psi_i}$, so $\ket{\Phi}$ is normalized and $\ket{\Ga} = \sqrt{1-\abs{\al_n}^2}\ket{\Phi} + \al_n \ket{\psi_n}$. By the induction hypothesis, we have a bound on $E(\phi)$:
\begin{align*}
E(\Phi) & \leq \Big(\sum_{i=1}^{n-1} \tfrac{\abs{\al_i}}{\sqrt{1-\abs{\al_n}^2}}\sqrt{E(\psi_i) + 1}\Big)^2 + n-3 \\
& = \frac{1}{1-\abs{\al_n}^2}\Big(\sum_{i=1}^{n-1} G_i\Big)^2 + n-3.
\end{align*}
Therefore,
\begin{align*}
G_{\Phi}^2 & := (1-\abs{\al_n}^2)(E(\Phi)+1) \\
& \leq \Big(\sum_{i=1}^{n-1} G_i\Big)^2 + (1-\abs{\al_n}^2)(n-2).
\end{align*}
Now, we use this bound to obtain a bound for $E(\Ga)$. Using Corollary \ref{cor:uppertwostate}, we have
\begin{align}
E(\Ga) & \leq G_n^2 + G_{\Phi}^2 + 2G_n G_\Phi \notag \\
& \leq G_n^2 + \Big(\sum_{i=1}^{n-1} G_i\Big)^2 + (1 -\abs{\al_n}^2)(n-2)  \notag \\ 
& \qquad + 2G_n\sqrt{\Big(\sum_{i=1}^{n-1} G_i\Big)^2 + (1-\abs{\al_n}^2)(n-2)}. \label{eqn:ega}
\end{align}
We work with the large square root term in line \eqref{eqn:ega}. Without loss of generality, order indices so that $\nobreak{\frac{G_n}{\abs{\al_n}^2} \leq \frac{G_i}{\abs{\al_i}^2}}$ for all $i < n$. Then $\abs{\al_i}^2 \leq G_i\abs{\al_n}^2/G_n$, so we have
\begin{align*}
\Big(\sum_{i=1}^{n-1} G_i\Big)^2 + & (1-\abs{\al_n}^2)(n-2) \\
& = \Big(\sum_{i=1}^{n-1} G_i\Big)^2 + (n-2) \sum_{i=1}^{n-1} \abs{\al_i}^2 \\
& \leq \Big(\sum_{i=1}^{n-1} G_i\Big)^2 + \frac{\abs{\al_n}^2(n-2)}{G_n}\sum_{i=1}^{n-1} G_i \\
& \leq \Big(\sum_{i=1}^{n-1} G_i + \frac{\abs{\al_n}^2(n-2)}{2G_n}\Big)^2. 
\end{align*}
Thus line \eqref{eqn:ega} reduces to 
\begin{align*}
E(\Ga) & \leq \Big(\sum_{i=1}^{n-1} G_i \Big)^2 + (1 -\abs{\al_n}^2)(n-2) + G_n^2 \notag \\ 
& \qquad + 2G_n\Big(\sum_{i=1}^{n-1} G_i + \frac{\abs{\al_n}^2(n-2)}{2G_n}\Big) \\
& = \Big(\sum_{i=1}^n G_i \Big)^2 + n-2.
\end{align*}
By induction, the result is proved.
\end{proof}

As the following two examples suggest, there are instances in which Theorem \ref{thm:upperinductionstate} is strictly better than Theorem \ref{thm:uppergenstate}, and vice versa. However, neither theorem is tight in general.

\begin{example}
\label{ex:uppergen}
If $\Ga$ is a trivial linear combination, say $\nobreak{\ket{\Ga} = \ket{\psi_1}}$, then choosing $p_1 = 1$, the bound from Theorem \ref{thm:uppergenstate} is tight:
\[
E(\Ga) \leq E(\psi_1).
\]
However the bound from Theorem \ref{thm:upperinductionstate} is not tight:
\[
E(\Ga) \leq E(\psi_1) + n-1.
\]
\end{example}

\begin{example}
\label{ex:uppergen2}
Define $\ket{\chi_0} := \frac{1}{\sqrt{d}}\sum_{i=1}^d \ket{ii}$, and define $\nobreak{\ket{\chi_j} := \ket{d+j,d+j}}$ for $j = 1,\ldots,n-1$. Also let $\om$ be an $n$-th primitive root of unity and for $k = 0,\ldots,n-1$, let
\[
\ket{\psi_k} := \frac{1}{\sqrt{n}} \sum_{j=0}^{n-1} \om^{jk} \ket{\chi_j}.
\]
Then $E(\psi_k) = \frac{1}{n} \log d + \log n$ for every $k$. Finally, let $\ket{\Ga} := \frac{1}{\sqrt{n}}\sum_{k=0}^{n-1} \ket{\psi_k} = \ket{\chi_0}$, so $E(\Ga) = \log d$. Choosing $p_i = 1/n$, the bound from Theorem \ref{thm:uppergenstate} is
\[
E(\Ga) \leq \log d + 2n \log n,
\]
while the bound from Theorem \ref{thm:upperinductionstate} is:
\[
E(\Ga) \leq \log d + n \log n + n-2.
\]
Letting $d \rightarrow \infty$, both bounds are asymptotically optimal; however, the bound from Theorem \ref{thm:upperinductionstate} is strictly better for $n > 2$. 
\end{example}

Next consider lower bounds. As with upper bounds, the starting point is result \cite[Theorem 4]{Gour07} for superpositions of two states:

\begin{theorem}
\label{thm:lowertwostate}
Let $\ket{\psi_1}$ and $\ket{\psi_2}$ be normalized, orthogonal bipartite states, and let $\ket{\Ga} = \al_1\ket{\psi_1} + \al_2\ket{\psi_2}$, also normalized. Then for any $0 \leq p \leq 1$,
\[
E(\Ga) \geq \frac{\abs{\al_1}^2E(\psi_1)}{1 + \frac{p\abs{\al_2}^2}{1-p}} - \frac{(1-p) E(\psi_2)}{p} - \frac{H(p)}{p},
\]
and similarly
\[
E(\Ga) \geq \frac{\abs{\al_2}^2E(\psi_2)}{1 + \frac{(1-p)\abs{\al_1}^2}{p}} - \frac{p E(\psi_1)}{1-p} - \frac{H(p)}{1-p}.
\]
\end{theorem}

Theorem \ref{thm:lowertwostate} can be generalized to superpositions of $k$ states.

\begin{theorem}
\label{thm:lowerbound}
Let $\{ \ket{\psi_i} \}_{i=1}^n$ be a set of normalized, orthogonal bipartite states, 
and let $\ket{\Ga} = \sum_i \al_i\ket{\psi_i}$, also normalized. Then for any $\{p_i\}$ such that $p_i \geq 0$ and $\sum_i p_i = 1$, and any $k \leq n$,
\[
E(\Ga) \geq \frac{\abs{\al_k}^2E(\psi_k)}{1 + p_k\sum_{i \neq k}\frac{\abs{\al_i}^2}{p_i}} - \sum_{i \neq k}\frac{p_i E(\psi_i)}{p_k} - \frac{H(\{p_i\})}{p_k}.
\]
\end{theorem}

\begin{proof}
Let $\ket{\chi_k} = \ket{\Ga}$, and suppose
\begin{equation}
\sum_{i\neq k} q_i \ketbra{\chi_i} + q_k\ketbra{\psi_k} = \sum_{i\neq k} p_i \ketbra{\psi_i} + p_k \ketbra{\Ga}.
\label{eqn:unittranslower}
\end{equation}
for some states $\{\ket{\chi_i}\}_{i \neq k}$. (Assuming that $\al_k \neq 0$, such $\ket{\chi_i}$ can be chosen. If $\al_k = 0$, then the bound in the statement of the theorem is trivial for that index $k$.) Then as in Theorem \ref{thm:uppergenstate},
\[
\sum_{i\neq k} q_i E(\chi_i) + q_k E(\psi_k) \leq \sum_{i\neq k} p_i E(\psi_i) + p_k E(\Ga) + H(\{p_i\}),
\]
and in particular,
\begin{equation}
p_k E(\Ga) \geq q_k E(\psi_k) - \sum_{i\neq k} p_i E(\psi_i) - H(\{p_i\}).
\label{eqn:gaboundlower}
\end{equation}
It remains to find the value of $q_k$ in terms of $\al_i$ and $p_i$. Let $(a_i)$ denote the last row of the unitary transformation $U$ relating $\sqrt{q_k} \ket{\psi_k}$ to $\sqrt{p_i} \ket{\psi_i}$ and $\sqrt{p_k} \ket{\Ga}$, so that 
\[
\sqrt{q_k} \ket{\psi_k} = \sum_{i \neq k} a_i\sqrt{p_i} \ket{\psi_i} +  a_k\sqrt{p_k} \ket{\Ga}.
\]
We also have $\ket{\Ga} = \sum_i \al_i\ket{\psi_i}$. So, matching the coefficients, we find that $a_k = \sqrt{q_k/p_k}/\al_k$, and also $\nobreak{\abs{a_i}^2 = \abs{\al_i/\al_k}^2q_k/p_i}$ for $i \neq k$. Now
\[
1 = \sum_i \abs{a_i}^2 = \frac{q_k}{\abs{\al_k}^2}\Big(\frac{1}{p_k} + \sum_{i\neq k} \frac{\abs{\al_i}^2}{p_i} \Big),
\]
so solving for $q_k$ implies that
\[
q_k = \frac{\abs{\al_k}^2}{\frac{1}{p_k} + \sum_{i\neq k} \frac{\abs{\al_i}^2}{p_i}},
\]
and the result follows by substituting $q_k$ into equation~\eqref{eqn:gaboundlower}. 
\end{proof}

While there are many examples for which the bound in Theorem \ref{thm:lowerbound} is worse than the trivial bound of $E(\Ga) \geq 0$, there are also cases where it is close to tight; \cite[Example~3]{Gour07} is one such example. More generally, the bound is tight whenever $\ket{\Ga} = \ket{\psi_k}$, so if $E(\psi_k) > 0$, then by continuity the bound is strictly positive for $\abs{\al_k}^2$ close to~$1$.
 
\section{Bounds on entanglement of subspaces}
\label{sec:spaces}

In this section we consider the maximum and minimum entanglement of a state in a subspace, as a function of the entanglement of basis states. More precisely, let $V$ be the span of $\{\ket{\psi_i}\}_{i=1}^n$, and define
\[
E_{\max}(V) := \max_{\ket{\phi} \in V} E(\phi), \qquad E_{\min}(V) := \min_{\ket{\phi} \in V} E(\phi).
\]
We aim to find upper bounds for $E_{\max}(V)$ and lower bounds for $E_{\min}(V)$ in terms of $\{E(\psi_i)\}$.

The results for $E_{\max}(V)$ are derived from maximizing entanglement of superpositions, using the results from the previous section. For a subspace of dimension $2$, the result is the following:

\begin{theorem}
\label{thm:uppertwospace}
Let $\ket{\psi_1}$ and $\ket{\psi_2}$ be normalized, orthogonal bipartite states, and let $V := \spn \{ \ket{\psi_1}, \ket{\psi_2} \}$. Then
\[
E_{\max}(V) \leq E(\psi_1) + E(\psi_2) + 2.
\]
\end{theorem}

\begin{proof}
If $\ket{\Ga}$ is an element of $V$, say $\ket{\Ga} = \sum_{i=1}^2 \al_i \ket{\psi_i}$, then Corollary \ref{cor:uppertwostate} gives a bound on $E(\Ga)$. Maximizing this bound over $\al_1$ and $\al_2$, the largest value occurs at
\begin{align*}
\abs{\al_1} & = \frac{\sqrt{E(\psi_1)+1}}{\sqrt{E(\psi_1)+E(\psi_2)+2}}, \\
\abs{\al_2} & = \frac{\sqrt{E(\psi_2)+1}}{\sqrt{E(\psi_1)+E(\psi_2)+2}}. 
\end{align*}
The bound on $E_{\max}(V)$ follows by substituting these choices of $\al_1$ and $\al_2$ into Corollary \ref{cor:uppertwostate}.
\end{proof}

Using Theorem \ref{thm:uppertwospace} recursively, we get a bound for subspaces of larger dimension:

\begin{theorem}
\label{thm:uppergenspace}
Let $\{ \ket{\psi_i} \}_{i=1}^n$ be a set of normalized, orthogonal bipartite states, and let $V := \spn \{\ket{\psi_i}\}_{i=1}^n$. Then 
\[
E_{\max}(V) \leq \sum_{i=1}^n E(\psi_i) + 2(n-1).
\]
\end{theorem}

\begin{proof}
This is an easy induction from Theorem \ref{thm:uppertwospace}, similar to the proof of Theorem \ref{thm:upperinductionstate}. 
\end{proof}

Alternatively, Theorem \ref{thm:uppergenspace} can be proven by taking the maximum of Theorem \ref{thm:upperinductionstate} over all $\al_i$. Note that Theorem~\ref{thm:uppergenspace} cannot be improved by taking the maximum of Theorem~\ref{thm:uppergenstate} over all $\al$. In fact, if $f(\al,p)$ denotes the bound in Theorem \ref{thm:uppergenstate}, namely 
\[
f(\al,p) := \left(\sum_i \frac{\abs{\al_i}^2}{p_i}\right)\Big(\sum_i p_i E(\psi_i) + H(\{p_i\})\Big),
\]
then 
\begin{equation}
\max_{\al} \min_p f(\al,p) = \sum_{i=1}^n E(\psi_i) + n \log n,
\label{eqn:falp}
\end{equation}
which, as a bound on $E_{\max}(V)$, is strictly worse than Theorem \ref{thm:uppergenspace} for $n > 2$. To see equation~\eqref{eqn:falp}, first note that choosing $p_i = 1/n$ for all $i$ makes $f(\al,p)$ independent of $\al$:
\[
\max_{\al} \min_p f(\al,p) \leq \max_{\al} f(\al,\tfrac{1}{n}) = \sum_{i=1}^n E(\psi_i) + n \log n.
\]
On the other hand, if we choose $\al = \hat{\al}$ such that 
\[
\abs{\hat{\al}_j}^2 := \frac{\log n+E(\psi_j)}{n\log n + \sum_i E(\psi_i)},
\]
it is possible to show that $\min_p f(\hat{\al},p)$ occurs at $p = 1/n$.

While the bound in Theorem \ref{thm:uppergenspace} does not appear to be tight, there is evidence to suggest that the bound is near optimal. 

\begin{example}
\label{ex:upperspace1}
Let $\ket{\chi}$ be a maximally entangled state in a $d$ by $d$ dimensional system, say $\ket{\chi} = \frac{1}{\sqrt{d}}\sum_{i=1}^d \ket{ii}$, and consider
\begin{align}
\ket{\psi_1} & := \sqrt{1-t}\ket{00} + \sqrt{t}\ket{\chi}, \label{eqn:psi1chi}\\
\ket{\psi_2} & := \sqrt{t}\ket{00} - \sqrt{1-t}\ket{\chi}, \label{eqn:psi2chi}
\end{align}
and $V = \spn\{\ket{\psi_1},\ket{\psi_2}\}$. Then $E(\psi_1) = t\log d + H(t)$ and $E(\psi_2) = (1-t)\log d + H(t)$; on the other hand, $\nobreak{E_{\max}(V) = \log (d+1)}$. Letting $d \rightarrow \infty$, we find that
\[
E_{\max}(V) \sim \sum_i E(\psi_i)
\]
for any choice of $t$. Therefore the bound in Theorem \ref{thm:uppergenspace} is asymptotically optimal for this family of subspaces.
\end{example}

\begin{example}
\label{ex:upperspace2}
Choose $\ket{\psi_i} := \ket{ii}$ and $V := \spn \{\ket{\psi_i}\}_{i=1}^n$, so that $E(\psi_i) = 0$ and $E_{\max}(V) = \log n$. It follows that if we have a bound on $E_{\max}(V)$ of the form 
\[
E_{\max}(V) \leq \sum_{i=1}^n E(\psi_i) + c,
\]
then $c \geq \log n$. 
\end{example}

We would also like to find lower bounds for $E_{\min}(V)$, the smallest entanglment in $V$. Unfortunately, no such bound exists, as the following two examples demonstrate. 

\begin{example}
\label{ex:lowerspace1}
Let $\ket{\chi}$ be any state with entanglement $c$, and then choose $\ket{\psi_1}$, $\ket{\psi_2}$ and $V$ as in equations \eqref{eqn:psi1chi} and \eqref{eqn:psi2chi}. Note that $\ket{00}$ is in $V$, so that $E_{\min}(V) = 0$. However,
\begin{align} \
E(\psi_1) & = tc + H(t); \label{eqn:lower1} \\
E(\psi_2) & = (1-t)c + H(t). \label{eqn:lower2}
\end{align}
It is not difficult to verify that for any nonnegative choices of $E(\psi_1)$ and $E(\psi_2)$, there are solutions $(t,c)$ to equations \eqref{eqn:lower1} and \eqref{eqn:lower2} in the ranges $0 \leq t \leq 1$ and $c \geq 0$. Therefore, for any two nonnegative numbers $E_1$ and $E_2$, there are choices of $\ket{\psi_1}$ and $\ket{\psi_2}$ such that $E(\psi_1) = E_1$, $E(\psi_2) = E_2$, and $\spn\{\ket{\psi_1},\ket{\psi_2}\}$ contains a separable state. It follows that there is no nontrivial lower bound on $E_{\min}(V)$ which is only a function of $E(\psi_1)$ and $E(\psi_2)$. In particular, both $E(\psi_1)$ and $E(\psi_2)$ can grow arbitrarily large while $E_{\min}(V)$ remains $0$. 
\end{example}


\begin{example}
\label{ex:lowerspace2}
Let $\ket{\chi_0} := \ket{00}$ and for $j = 1,\ldots,n-1$, let $\ket{\chi_j} := \frac{1}{\sqrt{d}}\sum_{i=dj+1}^{dj+d} \ket{ii}$. Also let $\om$ be an $n$-th primitive root of unity and for $k = 0,\ldots,n-1$ let
\[
\ket{\psi_k} := \frac{1}{\sqrt{n}} \sum_{j=0}^{n-1} \om^{jk} \ket{\chi_j}.
\]
Finally, let $V = \spn\{\ket{\psi_k}\}$. Clearly $V$ contains the separable state $\ket{\chi_0} = \ket{00}$, yet $E(\psi_k) = \frac{n-1}{n}\log d + \log n$ for every $k$. Letting $d \rightarrow \infty$, each $E(\psi_k)$ also goes to $\infty$, while $E_{\min}(V)$ remains $0$. We conclude that for any $n$, there is a subspace $V$ of dimension $n$ with a basis containing only elements of arbitrarily large entanglement, yet $V$ also contains a separable state.
\end{example}

\begin{example}
\label{ex:lowerspace3}
For $j = 0,\ldots,n-1$, let $\ket{\chi_j}$ be a maximally entangled state in a space of dimension $d^j$, such that the support subspaces of any $\ket{\chi_j}$ and $\ket{\chi_{j'}}$ are orthogonal. For example, take $\ket{\chi_0} := \ket{00}$, $\ket{\chi_1} = \frac{1}{\sqrt{d}}\sum_{i=1}^{d} \ket{ii}$, and so on. Then $E(\chi_j) = j \log d$. For $k = 0,\ldots,n-1$, define
\[
\ket{\psi_k} := \frac{n-2}{n}\ket{\chi_k} - \frac{2}{n} \sum_{j\neq k} \ket{\chi_j}.
\]
Finally, let $V = \spn\{\ket{\psi_k}\}$, so that $V$ contains the separable state $\ket{\chi_0}$ and $E_{\min}(V) = 0$. A simple calculation shows that for every $k$,
\[
E(\psi_k) = \frac{k(n-4)+2(n-1)}{n}\log d + c_n,
\]
where $c_n$ is a constant depending only on $n$. Letting $d \rightarrow \infty$, each $E(\psi_k)$ also goes to $\infty$, as does the difference between any $E(\psi_k)$ and $E(\psi_{k'})$ provided that $n>4$. We conclude that in contrast with the lower bound for $E(\Ga)$ in Theorem \ref{thm:lowerbound}, there is no lower bound for $E_{\min}(V)$ which grows linearly with the differences in $E(\psi_k)$. 
\end{example}

\section{Other measures of entanglement}
\label{sec:measures}

The technique used here to provide bounds on entanglement of superpositions can in fact be used with almost any measure of entanglement, not just the entropy of entanglement. In this section we outline the general technique and offer two measures of entanglement as specific examples: the tangle and the Schmidt rank.

Suppose $T$ is any measure of entanglement, and we wish to find an upper bound on $T(\Ga)$ as a function of $T(\psi_i)$, where $\ket{\Ga} = \sum_i \al_i\ket{\psi_i}$. To do this, let $\ket{\chi_1} = \ket{\Ga}$, and choose $\ket{\chi_2},\ldots, \ket{\chi_n}$ such that
\begin{equation}
\sum_i q_i \ketbra{\chi_i} = \sum_i p_i \ketbra{\psi_i};
\label{eqn:unittransgen}
\end{equation}
equation~\eqref{eqn:unittransgen} holds provided that $\{\sqrt{q_i}\ket{\chi_i}\}$ and $\{\sqrt{p_i}\ket{\psi_i}\}$ are related by a unitary $U$. Now suppose we can bound $T(\sum_i p_i \ketbra{\psi_i})$ from above and below by its components $T(\ketbra{\psi_i})$, say
\[
\sum_i f(p_i,T(\rho_i)) \leq T(\sum_i p_i \rho_i) \leq \sum_i g(p_i,T(\rho_i)),
\]
for some functions $f$ and $g$ with $f(p_i,T(\rho_i)) \geq 0$. Then from equation~\eqref{eqn:unittransgen},
\begin{equation}
f(q_1,T(\Ga)) \leq \sum_i g(p_i,T(\psi_i)).
\label{eqn:fgbound}
\end{equation}
The value of $q_1$ is the same as in Theorem \ref{thm:uppergenstate}, namely $
q_1 = (\sum_i \abs{\al_i}^2/p_i)^{-1}$. So, we obtain a bound from equation~\eqref{eqn:fgbound} substituting in $q_1$ and solving for $T(\Ga)$. 

Similarly, to obtain a lower bound, let $\ket{\chi_k} = \ket{\Ga}$, and choose $\ket{\chi_i}$ $(i \neq k)$ so that
\begin{equation}
\sum_{i\neq k} q_i \ketbra{\chi_i} + q_k\ketbra{\psi_k} = \sum_{i\neq k} p_i \ketbra{\psi_i} + p_k \ketbra{\Ga}.
\label{eqn:unittranslowergen}
\end{equation}
Then
\begin{equation}
g(p_k, T(\Ga)) \geq f(q_k,T(\psi_k)) - \sum_{i\neq k} g(p_i,T(\psi_i)).
\label{eqn:fgboundlower}
\end{equation}
As in Theorem \ref{thm:lowerbound}, $q_k$ is determined by $\al_i$ and $p_i$, namely $q_k = \abs{\al_k}^2(\frac{1}{p_k} + \sum_{i\neq k} \frac{\abs{\al_i}^2}{p_i})^{-1}$, which we substitute into equation~\eqref{eqn:unittranslowergen} and solve for $T(\Ga)$ to obtain a lower bound.

By way of example, let $T(\Ga)$ denote the \defn{tangle} of $\Ga$, which is defined as the linear entropy of the partial trace of $\Ga$. More precisely, let 
\[
S_L(\rho) := \frac{d}{d-1}\big(1- \tr(\rho^2)\big)
\]
be the (normalized) linear entropy of a $d \times d$ density matrix $\rho$. Then
\[
T(\Ga) := S_L(\tr_B \ketbra{\Ga}).
\]
Note that 
\[
S_L \Big(\sum_i p_i \rho_i \Big) = \sum_i p_i^2 S_L(\rho_i) + \sum_{i \neq j} p_ip_j \frac{d}{d-1}(1 - \tr(\rho_i\rho_j)),
\]
which can bounded above and below using positivity and the Cauchy-Schwarz inequality:
\[
0 \leq \tr(\rho_i\rho_j) \leq \sqrt{\tr(\rho_i^2)\tr(\rho_j^2)} \leq \sqrt{\tr(\rho_i^2)}.
\]
Therefore:
\[
S_L \Big(\sum_i p_i \rho_i \Big) \leq \sum_i p_i^2 S_L(\rho_i) + \frac{d}{d-1} \sum_i p_i(1-p_i),
\]
and
\[
\sum_i p_i^2 S_L(\rho_i) + \sum_{i \neq j} p_ip_j \frac{d}{d-1}\Big(\!1 - \sqrt{\tr(\rho_i^2)}\Big) \!\leq \!S_L \Big(\sum_i p_i \rho_i \Big).
\]
These bounds on linear entropy result in bounds on the tangle of $\Ga$ given in the following theorem.

\begin{theorem}
\label{thm:tanglebound}
Let $\{ \ket{\psi_i} \}_{i=1}^n$ be a set of normalized, orthogonal bipartite states, 
and let $\ket{\Ga} = \sum_i \al_i\ket{\psi_i}$, also normalized. Then for any $\{p_i\}$ such that $p_i \geq 0$ and $\sum_i p_i = 1$, we have
\[
T(\Ga) \leq \left(\sum_i \frac{\abs{\al_i}^2}{p_i}\right)\left(\sum_i p_i^2 T(\psi_i) + \frac{dp_i(1-p_i)}{d-1}\right),
\]
and for any $k \in \{1,\ldots,n\}$,
\[
T(\Ga) \geq \frac{q_k^2 T(\psi_k)}{p_k^2} - \sum_{i\neq k} \frac{p_i^2 T(\psi_i)}{p_k^2} - \sum_i \frac{dp_i(1-p_i)}{p_k^2(d-1)},
\]
where
\[
q_k = \frac{\abs{\al_k}^2}{\frac{1}{p_k} + \sum_{i \neq k} \frac{\abs{\al_i}^2}{p_i}}.
\]
\end{theorem}

Unfortunately, the bounds in Theorem \ref{thm:tanglebound} cannot be used to produce bounds on the largest tangle of a subspace, $T_{\max}(V) := \max_{\ket{\phi} \in V} E(\phi)$, or the smallest tangle, $T_{\min}(V) := \min_{\ket{\phi} \in V} E(\phi)$. To see that there is no bound on $T_{\min}(V)$, it suffices to note that the lower bound in Theorem \ref{thm:tanglebound} is never positive for $\abs{\al_i}^2 = 1/n$. Similarly, the upper bound in Theorem \ref{thm:tanglebound} is never less than $1$ for $\abs{\al_i}^2 = 1/n$, so there is no bound for $T_{\max}(V)$.

A measure of entanglement for which subspace lower bounds do exist is the Schmidt rank $r$, defined as follows:
\[
r(\Ga) := \rk(\tr_B \ketbra{\Ga}).
\]
For any two matrices $A$ and $B$, the rank of $A+B$ is bounded above by $\rk(A) + \rk(B)$. Using this fact and the decompositions of $\ket{\Ga} = \sum_i \al_i\ket{\psi_i}$ in equations \eqref{eqn:unittransgen} and \eqref{eqn:unittranslowergen}, we get the following result.

\begin{theorem}
\label{thm:schmidtbound}
Let $\{ \ket{\psi_i} \}_{i=1}^n$ be a set of normalized, orthogonal bipartite states, 
and let $\ket{\Ga} = \sum_i \al_i\ket{\psi_i}$, also normalized. Then 
\[
r(\Ga) \leq \sum_{i: \al_i \neq 0} r(\psi_i),
\]
and for any $k \in \{1,\ldots,n\}$ such that $\al_k \neq 0$,
\[
r(\Ga) \geq r(\psi_k) - \sum_{\substack{i\neq k \\ \al_i \neq 0}} r(\psi_i).
\]
\end{theorem}

Define $r_{\max}(V)$ and $r_{\min}(V)$ to be the largest and smallest Schmidt rank in subspace $V$ respectively.

\begin{corollary}
Let $\{ \ket{\psi_i} \}_{i=1}^n$ be a set of normalized, orthogonal bipartite states, and let $V := \spn \{\ket{\psi_i}\}_{i=1}^n$. Then 
\[
r_{\max}(V) \leq \sum_{i=1}^n r(\psi_i).
\]
Assume the indices $\{1,\ldots,n\}$ are chosen such that $r(\psi_1) \leq r(\psi_2) \leq \ldots \leq r(\psi_n)$. Then
\[
r_{\min}(V) \geq \min_{k=1}^n \left\{r(\psi_k) - \sum_{i=1}^{k-1} r(\psi_i) \right\}.
\]
\end{corollary}

In particular, if $V$ is the span of two orthogonal bipartite states $\ket{\psi_1}$ and $\ket{\psi_2}$, then
\[
r_{\max}(V) \leq r(\psi_1)+r(\psi_2),
\]
and
\[
r_{\min}(V) \geq \min\left\{r(\psi_1),r(\psi_2),\abs{r(\psi_1)-r(\psi_2)} \right\}.
\]

\section{Conclusions}
\label{sec:conclusions}

We have extended the results in~\cite{Gour07} to superpositions of several states rather than just two and used those results to find an upper bound on the maximum entanglement of a subspace. Examples indicate that for fixed subspace dimension, this bound is asymptotically optimal. We also show that it is impossible to find a lower bound on the minimum entanglement of a subspace in terms of the entanglement of the individual states in a basis. This is unfortunate, given that lower bounds might be used to establish additivity of entanglement of subspaces, or equivalently, the additivity of entanglement of formation. It also suggests that the minimum entanglement of a subspace depends strongly on the coherence among the states in the superposition rather than on the entanglement of the individual states in the basis. However, for at least one alternative measure of entanglement, namely the Schmidt rank, lower bounds for the entanglement of superpositions do imply the existence of lower bounds for the entanglement of subspaces. 

{\bf Acknowledgments:} This research has been supported by NSERC and MITACS. The authors would like to thank Barry Sanders for his helpful feedback and comments on this paper.


\begin{thebibliography}{1}

\bibitem{Gour07}
G. Gour, ``Entanglement of superpositions revisited,'' http://www.arxiv.org/abs/0704.1521, 2007.

\bibitem{Hor07} R. Horodecki, P. Horodecki, M. Horodecki and K. Horodecki,
``Quantum entanglement'', http://www.arxiv.org/abs/quant-ph/0702225, 2007.

\bibitem{Ple06} M. B. Plenio and S. Virmani, ``An introduction to entanglement measures'',
Quant. Inf. Comp. \textbf{7}, 1 (2007).

\bibitem{GourNolan07}
G. Gour and N.~R. Wallach, ``Entanglement of subspaces and error correcting codes'', Phys. Rev. A \textbf{76}, 042309, 2007.

\bibitem{Nolan02}
N. R. Wallach, ``An Unentangled Gleason's Theorem'', Contemporary Mathematics \textbf{305}, 291, 2002.

\bibitem{Hayden06} P. Hayden, D. Leung, and A. Winter, ``Aspects of generic entanglement,''
{\em Comm. Math. Phys.}, \textbf{265} (1):95--117, 2006.

\bibitem{Cubitt07} T.~S. Cubitt, A. Montanaro and A. Winter,
``On the dimension of subspaces with bounded Schmidt rank'', http://www.arxiv.org/abs/0706.0705, 2007.

\bibitem{Walgate07}
J. Walgate and A.~J. Scott, ``Generic local distinguishability and completely entangled subspaces,'' http://www.arxiv.org/abs/0709.4238, 2007.

\bibitem{Shor04}
P.~W. Shor, ``Equivalence of additivity questions in quantum information theory,'' {\em Comm. Math. Phys.}, \textbf{246} (3):453--472, 2004.

\bibitem{Linden06}
N. Linden, S. Popescu, and J.~A. Smolin, ``Entanglement of superpositions,'' {\em Physical Review Letters} \textbf{97} (10):100502, 2006.

\end{thebibliography}
\end{document}